# CLONING, EXPRESSION AND PURIFICATION OF THE GENERAL STRESS PROTEIN YhbO FROM *ESCHERICHIA COLI*

Jad Abdallah[1], Renee Kern[1], Abderrahim Malki, Viola Eckey, Gilbert Richarme*

Stress molecules, Institut Jacques Monod, Universite Paris 7, 2 place Jussieu, 75005 Paris, France

*Corresponding author

[1]The two first authors contributed equally to the work, and their name was arbitrarily choosen.

Tel  33 1 44 27 50 98

Fax 33 1 44 27 57 16

Email: richarme@ccr.jussieu.fr

Abbreviations: Hsp: heat shock protein°; amc°: 7-amino-4-methylcoumarin°; DTT°: dithiothreitol.

**ABSTRACT**

We cloned, expressed and purified the *Escherichia coli yhbO* gene product, which is homolog to the *Bacillus subtilis* general stress protein 18 (the *yfkM* gene product), the *Pyrococcus furiosus* intracellular protease PfpI, and the human Parkinson disease protein DJ-1. The gene coding for YhbO was generated by amplifying the *yhbO* gene from *E. coli* by polymerase chain reaction. It was inserted in the expression plasmid pET-21a, under the transcriptional control of the bacteriophage T7 promoter and *lac* operator. A BL21(DE3) *E. coli* strain transformed with the YhbO-expression vector *pET-21a-yhbO*, accumulates large amounts of a soluble protein of 20 kDa in SDS-PAGE that matches the expected YhbO molecular weight. YhbO was purified to homogeneity by HPLC DEAE ion exchange chromatography and hydroxylapatite chromatography and its identity was confirmed by N-terminal sequencing and mass spectrometry analysis. The native protein exists in monomeric, trimeric and hexameric forms.

**INTRODUCTION**

*Escherichia coli* YhbO is a member of the DJ-1/ThiJ/Pfp1 superfamily, which includes proteins with diverse functions, chaperones (E. coli Hsp31) (1-4), proteases (Pfp1 from *Pyrococcus furiosus* (5) and PHP1 from *Pyrococcus horikoshii* (6)), catalases, the Parkinson disease protein DJ-1 (7), and the ThiJ kinase involved in thiamine biosynthesis (8). The closest YhbO homologs are ThiJ in *E. coli*, YfkM and YraA in *B. subtilis* (involved in protection against environnemental stresses (9, 10)), the *Pyrococcus furiosus* protease 1 (5), and the Parkinson disease protein DJ-1 (for which different functions have been proposed, i.e. chaperone, peptidase, oxidative stress sensor) (7, 11, 13). The crystal structure of several members of the ThiJ superfamily has been solved. All members of the superfamily contain a similar domain with a nucleophilic elbow displaying an important cysteine, which is part, in Php1 and Hsp31, of a Cys, His, Glu/Asp catalytic triad responsible for their peptidase activity (3, 6). YhbO also possesses such a putative catalytic triade whereas DJ-1 contains instead a Cys, Glu diad (7).

Several members of this superfamily have been biochemically characterized. Hsp31 was characterized as a chaperone (3, 4) and a peptidase (14), and Pfp1 as a protease / peptidase active against gelatin and the fluorescent substrate AAF-amc (5). The biochemical characterization of DJ-1 led to contradictory results concerning putative chaperone, peptidase or redox activities (11, 13). The exact function of ThiJ in thiamine biosynthesis is not yet known.

Most of the members of the ThiJ superfamily function in cellular protection against environnemental stresses. The chaperone/peptidase Hsp31 is involved in thermal stress protection (2), YfkM and YraA in *Bacillus subtilis* protection against multiple stresses (9, 10), and DJ-1 in cellular protection against oxidative stress (13). *E. coli* YhbO is several fold overexpressed in stationary phase, during hyperosmotic stress or acid stress (15), and an *yhbO*-deficient strain is more sensitive than the parental strain to thermal, oxidative, hyperosmotic, pH, and UV stresses (manuscript in preparation), suggesting that YhbO is a general stress protein. In the present work, we report the cloning, expression and purification of *Escherichia coli* YhbO.

## MATERIALS AND METHODS

*Bacterial strains, plasmids, and growth conditons.*

The *E. coli* BL21 (DE3) strain (Novagen, USA) was used for the transformation of the synthetized DNA and the expression of YhbO. The gene coding for YhbO was generated by amplifying the *yhbO* gene from *E. coli* DNA by PCR using the forward primers (5 GGTGGTTGCTCTTCACATATGAGTAAGAAAATTGCC-3 ) containing a *Nde1* site and the reverse primer (5 GGTGGTCTGGGATCCTCATCAGGCAACAGACAGGCG-3 ) containing a *BamH1* site. The resulting product was digested with *Nde*1 and *BamH*1, ligated to the pET-21a (Novagen, USA) *Nde*1 and *BamH*1 backbone fragment, and transformed into strain BL21 (DE3).

*Bacterial extracts preparation.*

The YhbO overproducing strain (BL21 (DE3) pET-21a-*yhbO* was grown at 37¡C in 1 liter of Luria-Bertani medium (16) supplemented with ampicillin (50 g/ml) to an $OD_{600}$= 0.5. YhbO overexpression was induced with 1 mM IPTG and growth was continued until $OD_{600}$= 3. Cells were harvested by centrifugation at 4¡C. The cell pellet was resuspended in 10 ml of 30 mM Tris pH 8.0, 20 mM NaCl, 1 mM DTT, 0.5 mM EDTA, and bacteria were disrupted at 0¡C using a Branson sonicator (microtip, level 5, 5 x 30 seconds). This extract was centrifuged for 1 hour at 200,000 x g, and the supernatant was immediately loaded onto a HPLC DEAE column.

*HPLC DEAE ion exchange chromatography.*

YhbO was loaded on a TSK-DEAE-5PW (Tosohaas, Germany) equilibrated in 30 mM Tris pH 8.0, 20 mM NaCl, 1 mM DTT, 0.5 mM EDTA at 20¡C, and it was eluted (around 200 mM NaCl) with a linear gradient of 0.03 — 1 M NaCl in the same buffer. Fractions were analyzed by SDS-polyacrylamide gel electrophoresis and quantified by the Bradford assay.

*Hydroxyapatite chromatography.*

Purified fractions from the DEAE column were loaded on a 15 ml hydroxyapatite column (Bio-Gel HTP from BioRad), equilibrated in 20 mM Tris pH 7.5, 50 mM KCl, 1 mM DTT and eluted with a linear gradient of 0 - 100 mM sodium phosphate pH 7.5 in the same buffer. YhbO was dialyzed against 20 mM Tris pH 7.4, 50 mM KCL, 0.5 mM dithiothreitol and stored at —20¡C in buffer supplemented with 30% glycerol.

*Native molecular weight determination.*

The molecular weight of native YhbO was determined by filtration of the protein on a TSK-G4000-SW HPLC gel permeation column (Tosohaas, Germany). The column was equilibrated in 20 mM Tris pH 7.5, 100 mM NaCl, 1 mM DTT at 20¡C, loaded with 20 l of purified YhbO (2 mg/ml) and eluted at a flow rate of 0.5 ml/min. YhbO was detected by its absorbance at 280 nm. Alcohol dehydrogenase (molecular weight, 150,000), bovine serum albumin (molecular weight, 66,000), carbonic anhydrase (molecular weight, 29,000) and *E. coli* thioredoxin 1 (molecular weight, 12,000) were used as molecular weight standards, and were obtained from Sigma.

*Sodium dodecylsulfate-polyacrylamide gel electrophoresis.*

We perfored electrophoresis according to Laemmli, using 8-16% polyacrylamide gradient gels (Bio-Rad) with Coomassie blue staining. Broad range molecular weight markers were form Bio-Rad.

*Protein sequencing.*

N-terminal sequencing was performed by automated Edman degradation at the Laboratoire de Microsequen age des Prot ines, Pasteur Institute, Paris, with an Applied Biosystems 470 gas phase sequencer and an Applied Biosystems 1000S detector.

*Mass spectrometric identification of YhbO.*

Excised YhbO gel band was in-gel digested with mass spectrometry grade trypsin (Roche). Mass spectra were recorded in positive ion reflection mode of a matrix assisted laser desorption ionization-time of flight (MALDI-ToF) Voyager DE PRO (Applied Biosystems). Petide mass obtained were searched against *E. coli* database using the Mascot engine available online (www.matrixscience.com).

*Protease, peptidase and chaperone assays.*

Endopeptidase and aminopeptidase activities were assayed by monitoring the production of 7-amino-4-methylcoumarin (amc) from the fluoregenic aminoacids peptides Suc-LLVY-amc, Boc-LAR-amc, Ac-YVAD-amc, AAF-amc, ALK-amc, L-Ala-amc, L-Arg-amc, L-Asp-amc, L-Asn-amc, L-Gly-amc, L-Leu-amc, L-Lys-amc, L-Met-amc, L-Phe-amc, L-Pro-amc, L-Thr-amc, L-Tyr-amc, L-Val-amc, L-Ser-amc, L-Cys-amc, L-Glu-amc, L-His-amc as described in (14, 17). Carboxypeptidase activity was assayed by measuring the hydrolysis of Hippuryl-Phe or Hippuryl-Arg as described in (14): Gelatin-PAGE and casein-PAGE in-gel proteolysis were performed as described in (1, 5). Renaturation of urea-unfolded citrate synthase and protection of citrate synthase against aggregation at 43¡C were performed as described in (18).

*Reagents.*

The Klenow DNA polymerase was from Roche (Mannheim Germany), restriction enzymes were from Invitrogen, and the plasmid extraction kits was from Quiagen. Fluorogenic peptide substrates were from Bachem or Sigma, and all other chemicals were from Sigma and were reagent grade.

**RESULTS AND DISCUSSION**

*Construction of the YhbO-expression vector pET21a-yhbO.*

The *yhbO* gene from *E. coli* was amplified by PCR, with a *Nde1* and a *BamH1* site at the 5 and 3 end, respectively. The amplified product was about 550 bp (not shown), which is in accordance with the theoretical length of the *yhbO* gene (519 bp). It was cut wih *Nde1 and BamH1*, ligated to the *pET21a Nde1* and *BamH1* backbone fragment and transformed into strain BL21 (DE3) by electroporation. The constructed plasmid was verified by DNA sequencing (not shown).

*Expression of YhbO*

The BL21 (DE3) strain, transformed with the recombinant expression vector *pET-21a-yhbO*, and induced with 1 mM IPTG accumulates high amounts of a soluble protein migrating in SDS-PAGE with an apparent molecular weight of 20 kDa **(Figure 1, lane 2).** The overexpressed protein is not detected in an uninduced extract **(Figure 1, lane 1)**, and represents 28% in mass of the total proteins of the induced extract. This 20 kDa molecular weight matches the expected YhbO molecular weight of 19 kDa. Thus, YhbO migrates exclusively as a monomer in SDS polyacrylamide gels, in contrast with its *Pyrococcus furiosus* PfpI and *Pyrococcus Horikoshii* PhpI homolog which both display multmeric forms in SDS polyacrylamide gels (5, 6). YhbO was found in the soluble bacterial extract (200,000 x g supernatant), but not in the 200.000 x g pellet (not shown), suggesting that it neither forms inclusion bodies during overexpression nor does it colocalize with membrane fractions. The cytoplasmic localization of YhbO is in accordance its primary structure which does not contain any signal or membrane spanning sequence.

*YhbO purification*

YhbO was purified to homogeneity by two chromatographic steps on a HPLC DEAE ion exchange column and on a hydroxyapatite column **(Figure 1, lane 3 and 4, respectively).** Although YhbO appears pure after the first column, we performed a second chromatographic step in order to avoid a single-step purification procedure based on ion exchange chromatography. Since the protein was massively overexpressed under the T7 promoter (around 30% of total protein), overall purification was 3.5 fold, and yield was 77% (Table 1). The identity of YhbO was confirmed by N-terminal sequencing and mass spectrometry. The N-terminal sequence found was SKKIAVLI, which is identical to the YhbO N-terminal sequence without its N-terminal methionine (processing of the N-terminal methionine frequently occurs in bacteria). This sequence is not found in any of the other *E. coli* proteins. Mass spectrometry analysis identified unambiguously the purified protein as YhbO (**Figure 2A and 2B**).

*Quarternary structure*

The purified protein was analyzed by size exclusion chromatograhy on a SW G-4000 HPLC column equilibrated in 30 mM Tris pH 7.5, 100 mM NaCl, 1 mM DTT at 20¡C, as described in Materials and Methods. YhbO (20 l, 2 mg/ml) was loaded onto the column at a flow rate of 0.5 ml/min. It elutes under several peaks at 7.77 (expected molecular weight, 118 kDa), 8.88 (expected molecular weight, 58 kDa) and 10.23 min (expected molecular weight, 20 kDa) (**Figure 3A**). These peaks likely represent hexameric, trimeric and monomeric forms of YhbO, respectively **(Figure 3B).** Multimeric forms of PfpI-like proteases are commonly found (5, 6), and the crystal structure of PhpI shows a hexameric barrel-like oligomeric structure with the active sites sequestered inside the barrel (6).

*Absence of proteolytic, peptidolytic or chaperone activity in vitro*

We could not detect any proteolytic activity of YhbO, using an SDS/PAGE in-gel assay (used for the detection of the proteolytic activity of PfpI (5)) with gelatin or casein as substrate, or using the highly sensitive BODIPY-casein fluorescent test (11) (not shown). YhbO did not hydrolyze any of the 18 aminoacyl-amc substrates tested, in contrast with PepN (17) or Hsp31 (14). It did not hydrolyze either AlaAlaPhe-amc (the best substrate of *Pyrococcus furiosus* protease I (5)), the endopeptidase substrates acetyl-Ala-amc and Suc-AlaAlaAla-amc, and the carboxypeptidase substrates Hippuryl-Phe or Hippuryl-Arg (not shown).
Since YhbO presents some analogy with the Hsp31 chaperone/peptidase, we checked whether it displays chaperone properties. In contrast with Hsp31 (1), 20 M YhbO was unable to stimulate the renaturation of 0.2 M citrate synthase after denaturation in 8M urea (not shown), or to prevent citrate synthase aggregation upon thermal shock at 43¡C (not shown).

*Implications*

We cloned, overexpressed and purified to homogeneity *E. coli* YhbO. The overexpressed protein is found in the cytoplasm in a highly soluble form. We identified the purified protein by N-terminal sequencing and MALDI-ToF mass spectrometry. The expression and purification procedures in this study have provided a simple and efficient method to obtain pure *E. coli* YhbO in large quantities. The YhbO protein obtained will be used for further studies of its structure and function. YhbO exist as a monomer, trimer and hexamer like several archeae proteases (*Pyrococcus furiosus* proteaseI and *Pyrococcus horikoshii* proteaseI) However, we could not yet detect any proteolytic activity of YhbO using several protein or peptidic synthetic substrates, nor could we detect any chaperone activity, in contrast with Hsp31 which was recently characterized by us and others as a chaperone (1, 2) and a peptidase (14). Several members of the ThiJ domain superfamily still lack a precise biochemical characterization, like the Parkinson disease protein DJ-1, for which contradictory results have been reported concerning possible chaperone, peptidase and redox activities (7, 8, 11, 13). Thus, the physiological and biochemical characterization of the ThiJ superfamily stress proteins will still require intensive investigations.

**ACKNOWLEDGEMENTS.**

The authors wish to thank Dr. Jacques d'Alayer (Institut Pasteur, Paris) for N-terminal sequencing, Dr. Jean-Jacques Montagne for MALDI-ToF mass spectrometry experiments and A. Kropfinger for correction of the English language. This work was supported by grant CR 521090 from the DGA to G.R

**REFERENCES**

1  A. Malki, R. Kern, J. Abdallah, G. Richarme. Characterization of the *Escherichia coli* YedU protein as a molecular chaperone. Biochem. Biophys. Res. Commun. 301 (2003) 430-436.

2  M.S. Sastry,.K. Korotkov, Y. Brodsky, F. Baneyx. Hsp31, the *Escherichia coli* yedU gene product, is a molecular chaperone whose activity is inhibited by ATP at high temperatures. J. Biol. Chem. 277 (2002) 46026-46034.

3  P.M. Quigley, K. Korotkov, F. Baneyx, W. G. Hol. The 1.6-A crystal structure of the class of chaperones represented by *Escherichia coli* Hsp31 reveals a putative catalytic triad. Proc. Natl. Acad. Sci. U S A. 100 (2003) 3137-3142.

4 Y. Zhao, D. Liu, W.D. Kaluarachchi, H.D. Bellamy, M.A. White, R.O. Fox. The crystal structure of *Escherichia coli* heat shock protein YedU reveals three potential catalytic active sites. Protein Sci. 12 (2003) 2303-2311.

5 I.I. Blumentals, A.S. Robinson, R.M. Kelly. Characterization of sodium dodecyl sulfate-resistant proteolytic activity in the hyperthermophilic archaebacterium *Pyrococcus furiosus*. Appl. Environ. Microbiol. 56 (1990) 1992-1998.

6 X. Du, I.G. Choi, R. Kim, W. Wang, J. Jancarik, H. Yokota, S.H. Kim.  Crystal structure of an intracellular protease from *Pyrococcus horikoshii* at 2-A resolution. Proc. Natl. Acad. Sci. U S A. 97 (2000) 14079-14084.

7  S.J. Lee, S.J. Kim, I.K. Kim, J. Ko, C.S. Jeong, G.H. Kim, C. Park, S.O. Kang, P.G. Suh, H.S. Lee, S.S. Cha. Crystal structures of human DJ-1 and *Escherichia coli* Hsp31, which share an evolutionarily conserved domain. J. Biol. Chem. 278 (2003) 44552-44559.

8 S. Bandyopadhyay, M.R. Cookson. Evolutionary and functional relationships within the DJ1 superfamily. BMC Evol. Biol. 4 (2004) 6-12.

9 P.D. Thackray, A. Moir, SigM, an extracytoplasmic function sigma factor of *Bacillus subtilis*, is activated in response to cell wall antibiotics, ethanol, heat, acid, and superoxide stress. J. Bacteriol. (2003) 185:3491-3498.

10 A. Petersohn, H. Antelmann, U. Gerth, M. Hecker, Identification and transcriptional analysis of new members of the sigmaB regulon in *Bacillus subtilis*. Microbiology 145 (1999) 869-80.

11 J.A. Olzmann, K. Brown, K.D. Wilkinson, H.D. Rees, Q. Huai, H. Ke, A.I. Levey, L. Li, L.S. Chin, Familial Parkinson's disease-associated L166P mutation disrupts DJ-1 protein folding and function. J. Biol. Chem. 279 (2004) 8506-8515.

12 S.B. Halio, I.I. Blumentals, S.A. Short, B.M. Merrill, R.M. Kelly, Sequence, expression in *Escherichia coli*, and analysis of the gene encoding a novel intracellular protease (PfpI) from the hyperthermophilic archaeon Pyrococcus furiosus. J. Bacteriol. 178 (1996) 2605-2612.

13 S. Shendelman, A. Jonason, C. Martinat, T. Leete, A. Abeliovich. DJ-1 is a redox-dependent molecular chaperone that inhibits alpha-synuclein aggregate formation. PLoS Biol. 2 (2004) e362.

14 A. Malki, T. Caldas, J. Abdallah, R. Kern, S.J. Kim, S.S. Cha, H. Mori, G. Richarme, Peptidase activity of the Esch*erichia coli* Hsp31 chaperone. J. Biol. Chem. 280 (2005) 14420-14426.

15 H. Weber, T. Polen, J. Heuveling, V.F. Wendisch, R. Hengge, Genome-wide analysis of the general stress response network in *Escherichia coli*: sigmaS-dependent genes, promoters, and sigma factor selectivity. J. Bacteriol. 187 (2005) 1591-1603.

16 J.H. Miller, Experiments in Molecular Genetics, p. 439. Cold Spring Harbour Laboratory, Cold Spring Harbour, NY (1972).

17 D. Chandu, A. Kumar, D. Nandi, PepN, the major Suc-LLVY-AMC-hydrolyzing enzyme in *Escherichia coli*, displays functional similarity with downstream processing enzymes in Archaea and eukarya. Implications in cytosolic protein degradation. J. Biol. Chem. 278 (2003) 5548-5556.

18 T. Caldas, A. El Yaagoubi, G. Richarme, Chaperone properties of elongation factor EF-Tu. J. Biol. Chem. 273 (1998) 11478-11484.

**LEGENDS TO FIGURES**

**Figure 1. Purification of YhbO.** Protein samples were separated by sodium dodecylsulfate polyacrylamide gels (8-16% gradient), and stained with Coomassie brillant blue. *Lane 1*, crude extract from BL21(DE3), *pET-21a-yhbO* uninduced; *lane 2*, crude extract from BL21(DE3), *pET-21a-yhbO* induced with IPTG; *lane 3,* YhbO pool (8 μg) from the HPLC DEAE column, *lane 3,* YhbO pool (30 μg) from the hydroxyapatite column. The positions of the molecular weight markers (kDa) are indicated on the left.

**Figure 2. MALDI-ToF mass spectrum of the tryptic digest of *E. coli* YhbO.** A) MALDI-ToF spectrum. B) Matched peptides of YhbO.

**Figure 3. Native molecular weight determination.** Purified YhbO (20 l, 2 mg/ml) was loaded onto a TSK-G4000-SW HPLC gel permeation column as described under Experimental Procedures , and eluted at a rate of 0.5 ml/min. A) Chromatogram of the eluted fractions. Peaks represent the absorbance of YhbO at 280 nm. B) Elution time of the molecular weight markers as a function of the logarithm of their molecular weight: Alcohol dehydrogenase (150,000 Da), bovine serum albumin (66,000 Da), carbonic anhydrase (29,000 Da) and *E. coli* thioredoxin 1 (12,000 Da). The elution volume of the three YhbO peaks displayed in Figure 3A (at 7.77 min, 8.88 min and 10.23 min) are indicated by crosses marqued respectively H, T and M, corresponding to putative hexameric (around 12O kDa), trimeric (around 60 kDa) and monomeric (around 20 kDa) forms of YhbO.

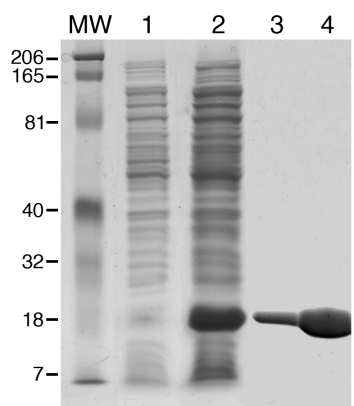

Figure 1

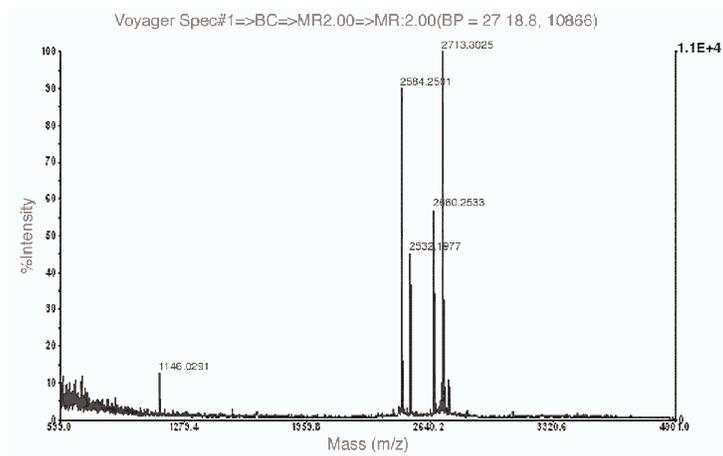

Figure 2A

```
MSKK**IAVLITDEFEDSEFTSPADEFR**KAGHEVITIEKQAGKTVKGKKGEASVTIDK**SIDE**      60

**VTPAEFDALLLPGGHSPDYLR**GDNRFVTFTRDFVNSGKPVFAICHGPQLLISADVIRGRK         120

LTAVKPIIIDVK**NAGAEFYDQEVVVDKDQLVTSRTPDDLPAFNR**EALRLLGA                172
```

Figure 2B. Matched peptide fragments in YhbO. The matched portions are shown in bold and underlined.

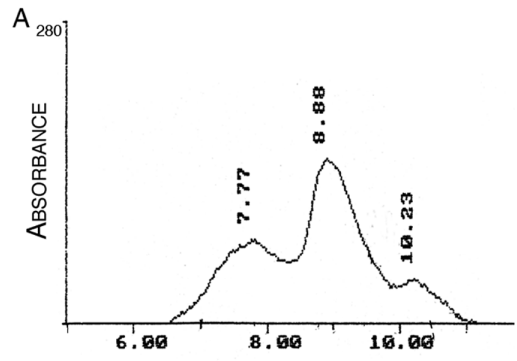

Figure 3A

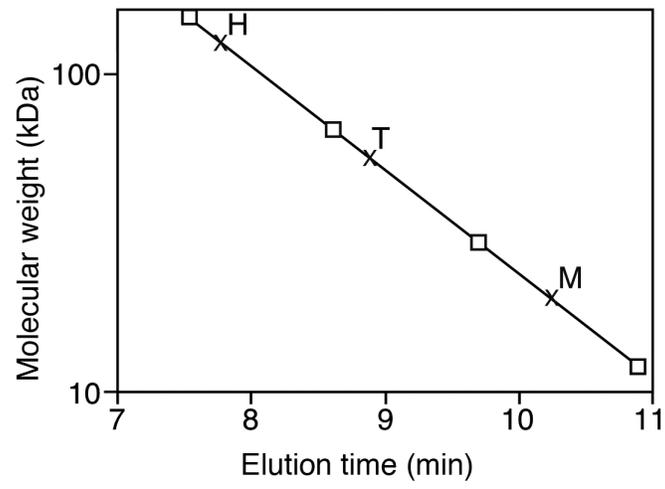

Figure 3B

| Purification step | Total protein (mg) | Total YhbO (mg) | Yield (%) | Overall purification (fold) |
|---|---|---|---|---|
| 200.000 x g supernatant | 220 | 62 | | |
| DEAE pool | 54 | 52 | 84 | 3.4 |
| Hydroxyapatite pool | 48 | 48 | 77 | 3.5 |

Table 1 : Purification of YhbO from *Escherichia coli*